\pdfoutput=1
\documentclass[12pt]{colt2022} 

\title[Properly learning decision trees in polynomial time?]{Open Problem: Properly learning decision trees in polynomial time?}
\usepackage{times}

\usepackage{changepage}

\coltauthor{%
 \Name{Guy Blanc} \Email{gblanc@stanford.edu}\\
 \addr Stanford 
 \AND
 \Name{Jane Lange} \Email{jlange@mit.edu}\\
 \addr MIT
 \AND
 \Name{Mingda Qiao} \Email{mqiao@stanford.edu}\\
 \addr Stanford
 \AND
  \Name{Li-Yang Tan} \Email{lytan@stanford.edu}\\
 \addr Stanford
}

\newcommand{\polylog}{\mathrm{polylog}}

\begin{document}

\maketitle

\begin{abstract}%
The authors recently gave an $n^{O(\log\log n)}$ time membership query algorithm for properly learning decision trees under the uniform distribution~\citep{BLQT21}.  The previous fastest algorithm for this problem ran in $n^{O(\log n)}$ time, a consequence of \cite{EH89}'s classic algorithm for the distribution-free setting.  In this article we highlight the natural open problem of obtaining a polynomial-time algorithm, discuss possible avenues towards obtaining it, and state intermediate milestones that we believe are of independent interest. 
\end{abstract}

\begin{keywords}%
 Decision trees, proper learning, analysis of Boolean functions 
\end{keywords}

\section{Introduction}

\newcommand{\zo}{\{0,1\}}
\newcommand{\bx}{\boldsymbol{x}}
\newcommand{\eps}{\varepsilon}
\newcommand{\poly}{\mathrm{poly}}

Decision trees are one of the most intensively studied concept classes in learning theory.  In this article we focus on the problem of {\sl properly} learning decision trees, where the learning algorithm is expected to return a hypothesis that is itself a decision tree.  Decision tree hypotheses are of particular interest because of their simple structure, and they are the canonical example of a highly interpretable model.  Indeed, it is natural to seek decision tree hypotheses even when learning concepts that are not themselves decision trees.  Algorithms such as ID3, CART, and C4.5 that construct decision tree representations of  datasets are among the most popular and empirically successful algorithms in everyday machine learning practice. 

We focus on the setting of PAC learning under the uniform distribution with membership queries, where the learner is given query access to an unknown size-$s$ decision tree target $f : \zo^n \to \zo$ and is expected to construct a decision tree hypothesis $h : \zo^n \to \zo$ satisfying $\Pr_{\bx\sim \zo^n}[f(\bx)\ne h(\bx)]\le \eps$, where $\bx \sim \zo^n$ is uniform random.  

The main open problem of this article is the following: 

\begin{openproblem}
\label{open:main}
Design a $\poly(n,s,1/\eps)$-time membership query algorithm for properly learning size-$s$ decision trees over $n$ variables to error $\eps$ under the uniform distribution. 
\end{openproblem}

Regarding the use of membership queries, it would of course be preferable if the algorithm does not require them and instead only uses uniform random labeled examples. However, there are two significant barriers to such an algorithm.  First, no such statistical query algorithm (SQ) exists: any SQ algorithm for learning size-$s$ decision trees has to make at least $n^{\Omega(\log s)}$ SQs~\citep{BFJKMR94}, and therefore run in at least that much time. Second, since every $k$-junta is a decision tree of size $2^k$, a $\poly(n,s)$ time algorithm for learning size-$s$ decision trees yields a polynomial-time algorithm for leaning $\log(n)$-juntas.  Designing such an algorithm that uses only random examples would represent a breakthrough on one of the central and longstanding open problems of learning theory~\citep{BL97}; indeed, it is reasonable to conjecture that there are no polynomial time algorithms for learning $k$-juntas from random examples for any $k = \omega_n(1)$. 

We also mention here the distinction between a {\sl strictly} proper versus {\sl weakly} proper algorithm: the former returns a size-$s$ decision tree hypothesis for a size-$s$ target, whereas the latter can return a decision tree of any size. Open Problem~\ref{open:main} is open even for weakly proper algorithms. 

\section{Background and current status} 

\cite{EH89} were the first to study the problem of properly learning decision trees.  They gave an $n^{O(\log s)}$-time weakly proper algorithm that works in the more general distribution-free setting and relies only on random examples.  Subsequently, two additional algorithms were designed in the uniform-distribution setting, both of which being substantially different from~\cite{EH89}'s and from each other.  First, \cite{MR02} gave an $n^{O(\log s)}$ time strongly proper algorithm. More recently, \cite{BLT-ITCS} designed a membership query algorithm that runs in $\poly(n)\cdot s^{O(\log s)}$ time.  For the standard setting where $s = \poly(n)$, all three algorithms run in quasipolynomial time, $n^{O(\log n)}$.  

Recent work of the authors gives a uniform-distribution membership query algorithm that runs in {\sl almost-polynomial} time, $n^{O(\log\log n)}$, bringing us a step closer to the resolution of Open Problem~\ref{open:main}.  The algorithm is strongly proper and additionally works in the agnostic setting~\citep{Hau92,KSS94}:

\begin{theorem}[\cite{BLQT21}] 
There is an algorithm which, given as input $\eps > 0$ and $s\in \mathbb{N}$, and query access to $f : \zo^n \to \zo$ that is promised to be $\mathrm{opt}_s$-close to a size-$s$ decision tree, runs in time 
\[ \tilde{O}(n^2) \cdot (s/\eps)^{O(\log((\log s)/\eps))} \] 
and outputs a size-$s$ decision tree $T$ that is w.h.p.~$(\mathrm{opt}_s + \eps)$-close to $f$.  If $f$ is monotone, the algorithm does not need membership queries and relies only on random examples. 
\end{theorem} 

Table~\ref{table} summarizes the performance guarantees of the algorithms discussed in this section.

\begin{table*}[ht!]
\begin{adjustwidth}{-3em}{}
\renewcommand{\arraystretch}{1.7}
\centering
\begin{tabular}{|c|c|c|c|c|}
\hline
  Reference  & Running time  & Hypothesis size   & ~~Access to target~~ &  Agnostic? \\ \hline
\cite{EH89} & $n^{O(\log s)}$ & $n^{O(\log s)}$ & Random examples & $\times$ \\ [.2em] \hline
 \cite{MR02} & $ n^{O(\log s)}$  & $s$ & Random examples & $\checkmark$ \\ [.2em]  \hline 
 \cite{BLT-ITCS} & $\poly(n)\cdot s^{O(\log s)}$ & $s^{O(\log s)}$ & Queries & $\times$ \\ [.2em] 
 \hline
 \cite{BLQT21} & ~~$\poly(n) \cdot s^{O(\log\log s)}$~~  & $s$ & Queries & $\checkmark$ \\ [.2em] \hline
\end{tabular}
\caption{Algorithms for properly learning size-$s$ decision trees.~\cite{EH89}'s algorithm works in the more general distribution-free setting, whereas all others work in the uniform-distribution setting.}  
\label{table}
\end{adjustwidth}
\end{table*}

\section{Possible approaches}

In this section we discuss how the techniques of~\cite{BLQT21} could lead to further progress on Open Problem~\ref{open:main}. 

\paragraph{Adaptive greediness.} The algorithm of \cite{BLT-ITCS} is a greedy algorithm: At each step, it places the variable with the largest influence, a natural measure of variable importance, as the root. This idea is extended in \cite{BLQT21}: Rather than greedily place down the variable with largest influence, that work proposes a brute force search of the top-$k$ most influential variables to find the best root. They are able to show that, setting $k = \mathrm{polylog}(s)$ and building a tree of depth $\log s$, the algorithm will build a high accuracy tree. The resulting runtime is $k^{O(\log s)} = s^{O(\log \log s)}$.

Perhaps $k$ need not be so big at every iteration? Can algorithms that choose a different amount of greediness (different $k$) at each step maintain accuracy while improving runtime? For example, it's not hard to show that an algorithm that uses $k = \mathrm{polylog}(s)$ for the first $\frac{\log s}{\log \log s}$ levels, and then $k = O(1)$ for the remaining $\log s - \frac{\log s}{\log \log s}$ would run in poly-time. Is such adaptive greediness sufficient to learn an accurate tree?

\paragraph{New splitting criteria.} The algorithm of~\cite{MR02} is based on a dynamic programming approach---to build an accurate decision tree for function $f$, we first solve the subproblems of building trees for restrictions of $f$ to at most $O(\log s)$ variables. A priori, there are $n^{O(\log s)}$ such restrictions, and this is the main bottleneck in the runtime. The algorithm of~\cite{BLQT21} can then be viewed as a refinement of this approach: Their structural result states that for each restricted function $f_{\pi}$, there exists a subset of $\polylog(s) \ll n$ variables that contains a near-optimal choice for the variable to be queried at the root. Furthermore, this subset can be identified by estimating the influence of each variable w.r.t.\ $f_{\pi}$. This allows us to reduce the number of subproblems to $(\polylog(s))^{O(\log s)} = s^{O(\log\log s)}$.

We note that even \emph{slightly} strengthening this structural result would immediately give a polynomial-time algorithm. Suppose that at each step, the algorithm tries to find the variables to be queried at the first $\ell$ levels of the tree (rather than only the root). Assuming that the number of such choices is at most $N_{\ell}$, the number of subproblems is at most $(N_{\ell}\cdot 2^{\ell})^{O(\log(s))/\ell}$, which is $\poly(s)$ if $N_{\ell} = 2^{O(\ell)}$. More concretely, can we pin down $\ell = \Theta(\log\log s)$ levels at once, while exploring only $\polylog(s)$ different choices at each step? If the influence itself is not enough for identifying a small subset of near-optimal choices, would a \emph{slightly} ``higher-order'' splitting criterion suffice?

\section{Intermediate milestones} 

In this section we state a couple of intermediate milestones towards Open Problem~\ref{open:main} that we believe are of independent interest.  

\paragraph{Monotone targets.} A function $f$ is \emph{monotone} if for any $x \prec x'$ in the partial order on $\zo^n$, we have $f(x) \le f(x')$. 
Monotone functions have some complexity restrictions that can allow for stronger learning algorithms.
For example, in the improper setting, polynomial-time learning algorithms exist for both monotone functions~\citep{OS07} and general functions~\citep{KM93},
but the learner for general functions requires membership queries
whereas the learner for monotone functions requires only random examples. 
A promising and independently interesting intermediate step towards Open Problem~\ref{open:main} is to design a polynomial-time proper learner under the assumption that the target function is monotone;
potentially even one that uses only random examples.

\paragraph{More expressive hypotheses.}  An algorithm for properly learning decision trees returns a decision tree hypothesis.  As an intermediate step towards such an algorithm, one could consider allowing the algorithm return more expressive hypotheses that nevertheless share the desirability of decision trees (e.g.~interpretability).  Two concrete examples are  branching programs---generalizations of decision trees where the underlying graph is a DAG---and generalized decision trees whose internal nodes branch on the outcomes of halfspaces instead of singleton variables.  Both these representations are easily seen to be exponentially more succinct than decision trees. In the context of boosting, branching program hypotheses have been shown to be enable exponential improvements over the decision tree hypotheses~\citep{MM02,LS05}.

\section*{Acknowledgements}

Guy and Li-Yang are supported by NSF CAREER Award 1942123. Mingda is supported by DOE Award DE-SC0019205 and ONR Young Investigator Award N00014-18-1-2295. Jane is supported by NSF Award CCF-2006664.

\bibliography{ref}

\end{document}